\documentclass{aa}  

\usepackage{graphicx}
\usepackage{txfonts}
\usepackage{natbib}
\usepackage{aalongtable}

\bibpunct{(}{)}{;}{a}{}{,}
\def\aap{A\&A}

\def\apj{ApJ}
\def\apjs{ApJS}
\def\apss{Ap\&SS}
\def\apjl{ApJ}

\def\mnras{MNRAS}

\begin{document}

\title{INTEGRAL-ISGRI observations of the Cyg\,OB2 region.}

\subtitle{Searching for hard X-ray point sources in a region containing several non-thermal emitting massive stars.}

\author{M. De Becker
        \inst{1}\fnmsep\thanks{Postdoctoral Researcher FNRS (Belgium).}
	\and
	G. Rauw
	\inst{1}\fnmsep\thanks{Research Associate FNRS (Belgium).}
	\and
	J.M. Pittard
	\inst{2}
	\and
	H. Sana
	\inst{3}
	\and
	I.R. Stevens
	\inst{4}
	\and
	G.E. Romero
	\inst{5,6}\fnmsep\thanks{Member of CONICET, Argentina.}
        }

\offprints{M. De Becker}

\institute{Institut d'Astrophysique et de G\'eophysique, Universit\'e de Li\`ege, 17, All\'ee du 6 Ao\^ut, B5c, B-4000 Sart Tilman, Belgium
	\and
	School of Physics and Astronomy, The University of Leeds, Woodhouse Lane, Leeds LS2 9JT, UK
	\and
	European Southern Observatory, Alonso de Cordova 3107, Vitacura, Casilla 19001, Santiago 19, Chile
	\and
	School of Physics and Astronomy, University of Birmingham, Edgbaston Birmingham B15 2TT, UK
	\and
	Facultad de Ciencias Astronom\'{\i}cas y Geof\'{\i}sicas, Universidad Nacional de La Plata, Paseo del Bosque, 1900 La Plata, Argentina
	\and
	Instituto Argentino de Radioastronom\'{\i}a, C.C.5, (1894) Villa Elisa, Buenos Aires, Argentina 
	}

\date{Received ; accepted }

\abstract  
   {}
   {We analyze \textit{INTEGRAL}-ISGRI data in order to probe the hard X-ray emission (above 20\,keV) from point sources in the Cyg\,OB2 region and to investigate the putative non-thermal high-energy emission from early-type stars (Wolf-Rayet and O-type stars). Among the targets located in the field of view, we focus on the still unidentified EGRET source 3EG\,2033+4118 that may be related to massive stars known to produce non-thermal emission in the radio domain, and on the wide colliding-wind binary WR\,140.}
   {Using a large set of data obtained with the IBIS-ISGRI imager onboard \textit{INTEGRAL}, we run the OSA software package in order to find point sources in the fully coded field of view of the instrument.}
   {Our data do not allow the detection of a lower-energy counterpart of 3EG\,J2033+4118 nor of any other new point sources in the field of view, and we derive upper limits on the high-energy flux for a few targets: 3EG\,J2033+4118, TeV\,J2032+4130, WR\,140, WR\,146 and WR\,147. The results are discussed in the context of the multiwavelength investigation of these objects.}
   {The upper limits derived are valuable constraints for models aimed at understanding the acceleration of particles in non-thermal emitting massive stars, and of the still unidentified very-high gamma-ray source TeV\,J2032+4130.}

\keywords{Stars: early-type -- Radiation mechanisms: non-thermal -- X-rays: stars -- Gamma rays: observations -- Acceleration of particles
               }

\maketitle               

\section{Introduction \label{intro}}

With the advent of the Energetic Gamma Ray Experiment Telescope (EGRET) onboard the Compton satellite, our vision of the gamma-ray sky improved significantly. However, among the 271 point sources listed in the third EGRET catalogue \citep{egretcat}, most are still unidentified. Many of these high-energy sources can be associated to supernova remnants, active galactic nuclei, pulsars and High-Mass or Low-Mass X-Ray Binaries (respectively HMXRB and LMXRB), but it has also been shown that a few may be coincident with early-type stars \citep[see e.g.][]{romero-egret}. A good example is the EGRET source 3EG\,J2033+4118 that is located in Cyg\,OB2, one of the richest OB associations of the Galaxy \citep{knodlseder}.\\

Cyg\,OB2 is also interesting in the sense that it harbours 3 O-type stars (Cyg\,OB2\,\#5, \#8A and \#9) known to produce non-thermal radio emission, revealing therefore that these stars are able to accelerate electrons up to relativistic energies. The existence of such a population of relativistic particles opens up the possibility that other non-thermal emission processes are at work in the high-energy domain. For this reason, non-thermal radio emitting early-type stars are considered as candidates for the emission of non-thermal radiation in the hard X-rays and in the $\gamma$-rays \citep[see e.g.][]{debecker-thesis}. The putative contribution of some of these early-type stars to the $\gamma$-ray source 3EG\,J2033+4118 has already been discussed by \citet{benaglia-5} and \citet{debecker-hk}. In addition, a few long-period Wolf-Rayet binaries (WR\,140, WR\,146, and WR\,147) located close to Cyg\,OB2 are also classified as non-thermal radio emitters, and may therefore be non-thermal high-energy sources. In the context of the so-called `standard' model for the non-thermal emission from massive stars, the electrons are accelerated through the Diffusive Shock Acceleration mechanism (DSA, \citealt{pittard-DSA}) by hydrodynamic shocks produced by colliding winds in binary systems. The high-energy emission is expected to arise from the inverse Compton (IC) scattering of UV photons emitted by the stars of the system, even though hadronic processes such as neutral pion decay may also contribute to the $\gamma$-rays. However, there are many uncertain parameters in current models which require observations to further constrain the nature of the acceleration process and the efficiency of leptonic and hadronic emission processes. Recent observations with {\it XMM-Newton} (between 0.5 and 10.0\,keV) of non-thermal radio emitting early-type stars failed to detect unambigously a non-thermal soft (i.e. below 10\,keV) X-ray emission component \citep{rauw-9sgr,debecker-168112,debecker-167971,debecker-cyg8a}. This non-detection may be explained by the limited availability of UV photons for IC scattering in binary systems characterized by somewhat large separations, and mostly by the fact that the faint putative non-thermal emission may be overwhelmed by the thermal emission from these systems that is much stronger in soft X-rays. For this reason, an investigation of the higher energy domain -- where there is no more thermal emission from colliding-winds -- is strongly needed.

The existence of the very high-energy $\gamma$-ray source TeV\,J2032+4130 \citep{TeV,TeV2} in the direction of Cyg\,OB2 -- discovered using Cherenkov telescopes -- should also be considered. The nature of this source is still unknown, even though it has been proposed that it may be related to the rich population of massive stars in the Cyg\,OB2 region \citep{buttcwe-tev,torres-assoc}. The putative relation of TeV\,J2032+4130 with 3EG\,J2033+4118 is also worth considering, even though their error boxes are only marginally consistent.\\

On the basis of a large set of data obtained with the International Gamma-Ray Laboratory (\textit{INTEGRAL}), \citet{debecker-thesis} searched for the presence of high-energy sources related to the massive star population of Cyg\,OB2, but failed to detect the targets mentioned above. In this paper, we discuss a larger set of {\it INTEGRAL}-ISGRI data in order to investigate the hard X-ray emission from the Cyg\,OB2 region, with the purpose to constrain the flux of the targets mentioned above in the ISGRI bandpass, i.e. between about 20\,keV and 1\,MeV. The results are also considered in the context of the multiwavelength investigation of the non-thermal emission of radiation from astrophysical sources.

\section{Observations and data processing \label{data}}

Time was granted (PI: G. Rauw) to observe the Cyg\,OB2 region with the IBIS imager \citep{ibis} onboard \textit{INTEGRAL} during revolutions 0080 (Announcement of Opportunity number 1, AO1), and in revolutions 0191, 0210, 0211, 0212, 0213, 0214, 0215, 0216, 0218, 0251, 0252, 0253, 0254 and 0255 (AO2), and these data were analyzed by \citet{debecker-thesis}. In addition, the same field was observed on the request of many teams, including in the context of the Guaranteed Time.

\begin{figure*}[ht]
\centering
\includegraphics[width=12cm]{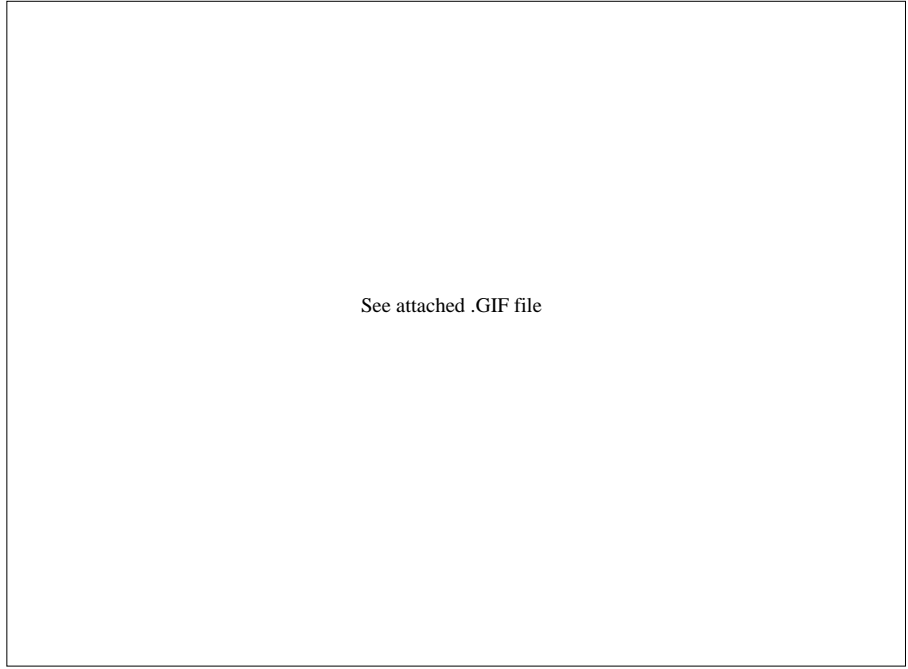}
\caption{IBIS-ISGRI mosaic image constructed on the basis of the 715 science windows (FCFOV) between 20 and 60\,keV. The point sources detected in the field of view are individually pointed out. The large structures centered on Cyg\,X-1 are artificial. The coordinate grid specifies the right ascension and the declination.\label{mosaic}}
\end{figure*}

The data set (observing group) discussed in this paper is constituted of all public science windows (up to revolution 340) where the position of 3EG\,J2033+4118 appears in the Fully Coded Field of View (FCFOV). We did not consider the Partially Coded Field of View (PCFOV) in order to reduce the impact of noise in our data set, therefore optimizing the efficiency of the source detection procedure. This observing group contains 715 science windows of about 50 minutes each on average, leading to a total effective observation time of about 2120\,ks. We applied the standard ISGRI\footnote{As we were interested mainly in a priori rather soft faint sources, we did not consider PICsIT data in our analysis.} data analysis procedure using the OSA software (v6.0) provided by the Integral Science Data Center (ISDC, \citealt{isdc}) in order to build a mosaic image and to detect sources. We distributed the events into three energy bands: (1) 20 -- 60\,keV, (2) 60 -- 100\,keV, and (3) 100 -- 1000\,keV. The mosaic image obtained between 20 and 60\,keV is shown in Fig.\,\ref{mosaic}.

\begin{table*}[ht]
\begin{center}
\caption{Input catalogue used for the high-level analysis of the FCFOV around the expected position of 3EG\,J2033+4118. The sources are sorted by decreasing detection significance level in the first energy band. The last three columns provide the significance of the detection of the sources respectively in the three energy bands selected for the data analysis. In the last three columns, `--' indicates a non detection. References to previous {\it INTEGRAL} observations of most of these sources are given below.\label{initcat}}
\vspace*{3mm}
\begin{tabular}{cllllllll}
\hline
Source & Nature & Ref &  $\alpha$ (J2000) & $\delta$ (J2000) & Status & $\sigma_1$ & $\sigma_2$ & $\sigma_3$\\
\hline
Cyg\,X-1 & HMXRB & 2 & 19h\,58m\,21.7s & +35$^\circ$\,12'\,06'' & Detected & 4204 & 1492 & 480 \\
Cyg\,X-3 & Microquasar & 5 & 20h\,32m\,26.6s & +40$^\circ$\,57'\,09'' & Detected & 1252 & 110 & 20 \\
EXO\,2030+375 & Be/X-ray binary & 4,6 & 20h\,32m\,15.2s & +37$^\circ$\,38'\,15'' & Detected & 177 & 21 & -- \\
Cyg\,X-2 & LMXRB & 7 & 21h\,44m\,41.2s & +38$^\circ$\,19'\,18'' & Detected & 65 & -- & -- \\
SAX\,J2103.5+4545 & HMXRB & 8 & 21h\,03m\,33.0s & +45$^\circ$\,45'\,00'' & Detected & 62 & 7 & -- \\
KS\,1947+300 & Be/X-ray binary & 9 & 19h\,49m\,35.6s & +30$^\circ$\,12'\,31'' & Detected & 47 & 10 & -- \\
QSO\,B1957+405 & Seyfert 1 galaxy & 1 & 19h\,59m\,28.4s & +40$^\circ$\,44'\,02'' & Detected & 30 & 11 & -- \\
IGR\,J21247+5058 & gamma-ray source & 10 & 21h\,24m\,42.0s & +50$^\circ$\,59'\,00'' & Detected & 22 & 7 & -- \\
SS\,Cyg & Dwarf nova & 3 & 21h\,42m\,48.0s & +43$^\circ$\,34'\,36'' & Detected & 12 & -- & -- \\
IGR\,J21335+5105 & gamma-ray source & -- & 21h\,33m\,50.1s & +51$^\circ$\,09'\,22'' & Not detected & -- & -- & -- \\
3EG\,J2033+4118 & gamma-ray source & -- & 20h\,33m\,36.0s & +41$^\circ$\,19'\,00'' & Not detected & -- & -- & -- \\
TeV\,J2032+4130 & gamma-ray source & -- & 20h\,32m\,07.0s & +41$^\circ$\,30'\,30'' & Not detected & -- & -- & -- \\
WR\,140 & Wolf-Rayet binary & -- & 20h\,20m\,28.0s & +43$^\circ$\,51'\,16'' & Not detected & -- & -- & -- \\
WR\,146 & Wolf-Rayet binary & -- & 20h\,35m\,45.1s & +41$^\circ$\,22'\,44'' & Not detected & -- & -- & -- \\
WR\,147 & Wolf-Rayet binary & -- & 20h\,36m\,43.7s & +40$^\circ$\,21'\,07'' & Not detected & -- & -- & -- \\
\hline
\end{tabular}
\end{center}
(1) \citet{bassani-qso}; (2) \citet{bazzano-cygX1}; (3) \citet{softcata-isgri}; (4) \citet{camero-exo}; (5) \citet{goldoni-cygX3}; (6) \citet{kuznetsov-exo}; (7) \citet{natalucci-cygX2}; (8) \citet{sidoli-sax}; (9) \citet{tsygankov-ks}; (10) \citet{walter-igr}
\end{table*}

The detection threshold was fixed at 3\,$\sigma$. We forced the detection procedure to search only for the sources included in an input catalogue containing a restricted number of potential sources. This approach is useful when dealing with noisy science windows and prevents detection of artificial sources due to ghosts of the bright sources Cyg\,X-1 and Cyg\,X-3. The input catalogue (see Table\,\ref{initcat}) contains 11 previously known sources\footnote{These sources are those that were already detected by \citet{debecker-thesis}, along with IGR\,J21223+5105 and SS\,Cyg that were apparent point sources quoted as `s1' and `s2' by \citet{debecker-thesis} although they were not detected with a significance above 3\,$\sigma$.}. In addition, we included 3EG\,J2033+4118, the three Wolf-Rayet stars discussed in Sect.\,\ref{intro} and TeV\,J2032+4130. The coordinates we used for the TeV source are those proposed by \citet{TeV2} whose analysis of Whipple Observatory data led to a shift of about 9 arcmin with respect to the position initially given by the High Energy Gamma Ray Astronomy consortium \citep{TeV}.

\section{Results and discussion}

\begin{figure}[ht]
\centering
\includegraphics[width=8.5cm]{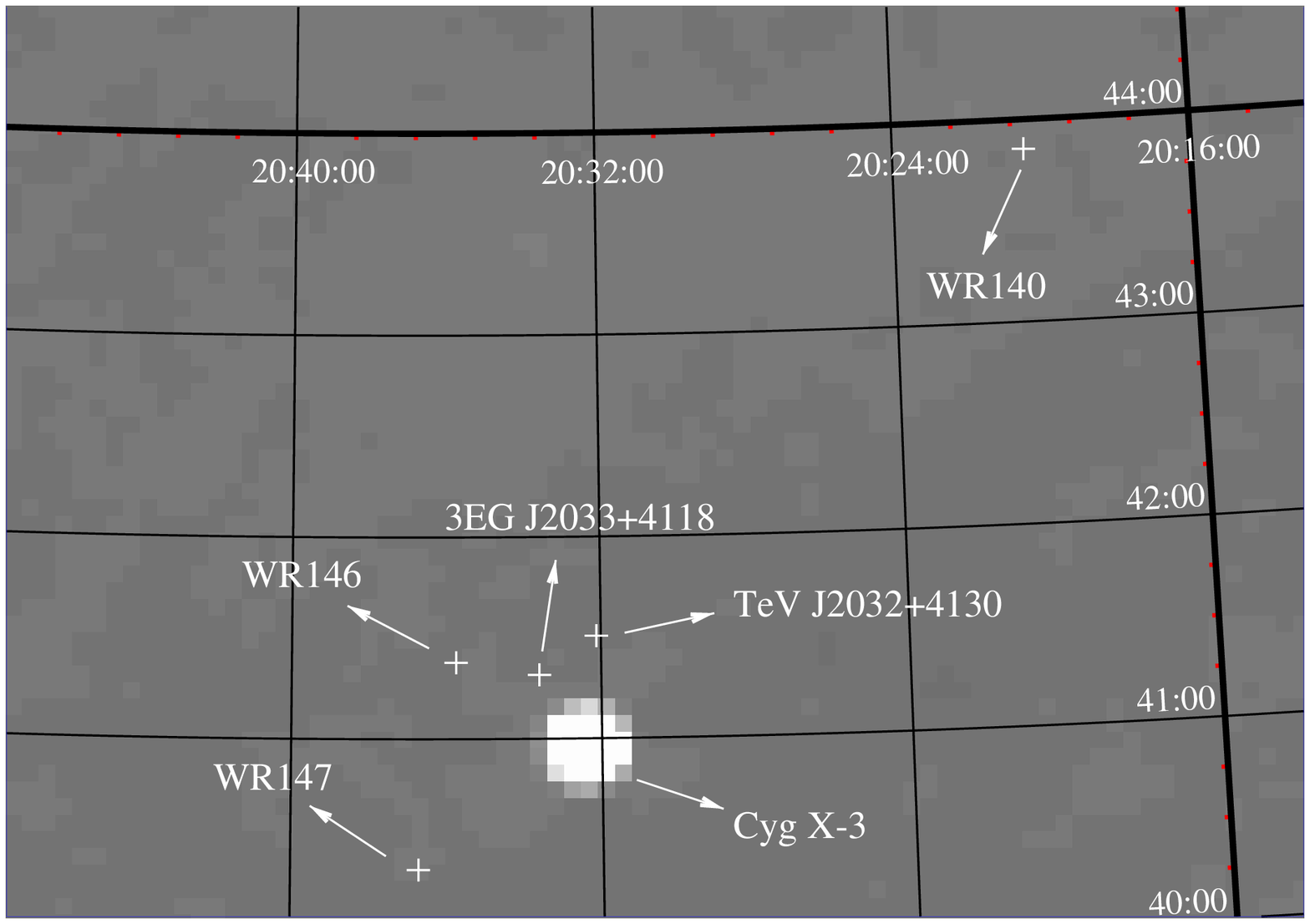}
\caption{Fraction of the IBIS-ISGRI mosaic image between 20 and 60\,keV (see Fig\,\ref{mosaic}). The position of our main targets is specified. The bright source on this image is Cyg X-3. The coordinate grid specifies the right ascension and the declination.\label{mosazoom}}
\end{figure}

The analysis of our data did not allow us to detect any of the targets that motivated this study. A closer view of the intensity map between 20 and 60\,keV -- where the expected position of the undetected sources is indicated -- is shown in Fig.\,\ref{mosazoom}. A summary of the source detection results is given in Table\,\ref{initcat}, where the significance of the detection is specified for the three energy bands. The main reason for this non-detection could be that we are dealing with a priori faint high-energy sources located in a field populated by several bright sources, among which the very bright High Mass X-Ray Binary Cyg\,X-1 is dominant. The efficiency of the detection procedure is indeed hampered by the artificial large scale structures due to the presence of Cyg\,X-1\footnote{Every science window image in the FCFOV was carefully inspected in order to reject those presenting the strongest artefacts (chessboard structures, bright ghosts, noisy structures). However, this selection (up to 60\,\% of the data rejected) did not reduce the large scale artificial structures centered around Cyg\,X-1 that are still seen in the mosaic image obtained with the data set discussed in this paper (see Fig.\,\ref{mosaic}).} that heterogeneously increase the level of the background\footnote{Without these structures and artefacts, the background would be dominated by a diffuse extragalactic emission in the soft part of the ISGRI bandpass, and by the internal background due to the de-excitation of nuclei produced by spallation reactions of cosmic rays on the instrument material at higher energies \citep{lebrun-isgri}.}.

On the basis of the mosaic image of the FCFOV, we estimated upper limits on the high-energy flux in the three energy bands specified in Sect.\,\ref{data} using the {\tt mosaic\_spec} task. We first fitted a Gaussian (half width at half maximum fixed at 6 arcmin) at the expected position of each target in order to measure a count rate. We repeated the same procedure by shifting the position of the Gaussian by 12 arcmin, i.e. the expected angular resolution of IBIS. Four shifts were applied respectively in the North, South, East and West directions in order to check for the consistency of the results of the fit across the typical positional error box of IBIS. Variations are indeed expected if the background is not homogeneous in that part of the image.

\subsection{3EG\,J2033+4118}

The upper limits on the count rates obtained in the three energy bands for 3EG\,J2033+4118 are very consistent across the typical error box of IBIS. This suggests that the background is rather homogeneous in this part of the image even though the target position is very close to that of Cyg\,X-3.

\begin{table}[h]
\centering
\caption{Upper limits at 3\,$\sigma$ on the count rate and on the fluxes in three energy bands for 3EG\,J2033+4118 and WR\,140.\label{upperlimits}}
\vspace*{3mm}
\begin{tabular}{llll}
\hline
 & \multicolumn{3}{c}{Upper limits (3\,$\sigma$)} \\
\cline{2-4}
Energy band & Count rate & Photon flux & Flux\\
(keV) & (cts\,s$^{-1}$) & (ph\,cm$^{-2}$\,s$^{-1}$) & (erg\,cm$^{-2}$\,s$^{-1}$) \\
\hline
\multicolumn{3}{l}{3EG\,J2033+4118} & \\
\hline
20 -- 60 & 0.087 & 1.1\,$\times$\,10$^{-4}$ & 6.1\,$\times$\,10$^{-12}$ \\
60 -- 100 & 0.051 & 3.4\,$\times$\,10$^{-5}$ & 4.2\,$\times$\,10$^{-12}$ \\
100 -- 1000 & 0.069 & 7.9\,$\times$\,10$^{-5}$ & 4.0\,$\times$\,10$^{-11}$ \\
\hline
\multicolumn{3}{l}{WR\,140} & \\
\hline
20 -- 60 & 0.090 & 1.1\,$\times$\,10$^{-4}$ & 6.3\,$\times$\,10$^{-12}$ \\
60 -- 100 & 0.054 & 3.5\,$\times$\,10$^{-5}$ & 4.4\,$\times$\,10$^{-12}$\\
100 -- 1000 & 0.072 & 8.2\,$\times$\,10$^{-5}$ & 4.2\,$\times$\,10$^{-11}$\\
\hline
\end{tabular}
\end{table}

As a second step, we converted these upper limits on the count rates into fluxes expressed in erg\,cm$^{-2}$\,s$^{-1}$ and in ph\,cm$^{-2}$\,s$^{-1}$. It is therefore necessary at this stage to make an assumption on the model of the high-energy emission in the energy bands used in our ISGRI data analysis. Considering that the high-energy emission is produced by IC scattering, we may expect an emission spectrum in the form of a power law with a photon index equal to 1.5 (this value is expected for IC emission produced by a population of relativistic electrons accelerated by the DSA mechanism in the presence of strong shocks, see e.g. \citealt{debecker-thesis}). We built a synthetic model by folding such a power law affected by an arbitrary normalization parameter with the response matrices of ISGRI (response matrix and ancillary response) using the {\sc xspec} software. We then scaled the normalization parameter of the model in order to match the count rate of the fake spectrum with the upper limit on the count rate in the complete ISGRI bandpass. The flux was then estimated on the basis of this scaled synthetic model in each energy band. The upper limits on the fluxes are given in Table\,\ref{upperlimits}.

As mentioned in the introduction, this unidentified $\gamma$-ray source may be related to the population of O-type stars located in Cyg\,OB2. In such a scenario, the high-energy emission may be explained by the combined IC emission from the three non-thermal radio emitting stars Cyg\,OB2\,\#5, \#8A and \#9. In order to disentangle the putative high-energy contributions from these objects, we need to determine accurately their stellar, wind, and orbital parameters. The best known system is undoubtedly Cyg\,OB2\,\#8A whose orbital solution \citep{debecker-cyg8asol} has been further validated by radio and X-ray observations revealing strong phase-locked variations with the same ephemeris \citep{blomme-jenam,debecker-cyg8a}. Cyg\,OB2\,\#5 is known to be a triple system \citep{contreras-cyg5} but no orbital solution exists for the third component. For Cyg\,OB2\,\#9, the existence of a companion has not yet been revealed even though strong variations of the radio flux are in agreement with a long period binary scenario \citep{vanloo-thesis}. An optical campaign is currently under way aiming at the investigation of the multiplicity of this latter target. Provided detailed information are gathered on these objects, state-of-the-art models \citep[see e.g.][]{pittard-DSA} may be applied in order to estimate the respective contributions to the expected non-thermal high-energy emission from these O-type stars. This constitutes the obvious next step in the study of the non-thermal phenomena related to these early-type stars.\\

It should be noted that the power law model generally used to reproduce the expected non-thermal high-energy emission from colliding-wind binaries does not necessarily hold in the EGRET bandpass. On the one hand, the emission process at work above 100\,MeV may be different form IC scattering, as it may be due to a hadronic process such as neutral pion decay. On the other hand, assuming that the $\gamma$-rays detected by EGRET come from IC emission, the index of the power law at these energies may be very different from that characterizing the spectrum at a few tens of keV. The extrapolation to keV energies of the EGRET emission level measured above 100\,MeV is therefore not expected to carry any relevant physical meaning.

\subsection{WR stars}

We also estimated the upper limits on the flux following the same procedure as above in the case of the WR stars included in our catalogue. The first target worth considering here is the long period binary WR\,140 (WC7 + O4-5). According to the ephemeris published by \citet{marchenko-wr140}, our data set covers orbital phases between 0.23 and 0.56, even though most of the science windows ($\sim$\,65\%) have been obtained between phases 0.41 and 0.48. For WR\,140 \footnote{We note that the position of WR\,140 is coincident with the error box of the EGRET source 3EG\,J2022+4317 \citep{romero-egret,benaglia-romero}.}, the upper limits are slightly different to those obtained for 3EG\,J2033+4118 (see Table\,\ref{upperlimits}). We note that these values are indeed larger than the level of high energy emission predicted by several colliding-wind binary models developed by \citet{pittard-DSA}\footnote{The approach developed by \citet{pittard-DSA} consists in fitting models to radio data in order to determine the population and spatial distribution of relativistic electrons without an a priori knowledge of the magnetic field strength.} for WR\,140. However, the fluxes predicted by some models used by \citet{pittard-DSA} are larger than the upper limits deduced from the data. In particular, the predicted flux in the 60-100 keV energy band of model H of \citet{pittard-DSA} is 4.2\,$\times$\,10$^{-5}$\,ph\,cm$^{-2}$\,s$^{-1}$ (5.6\,$\times$\,10$^{-12}$\,erg\,cm$^{-2}$\,s$^{-1}$). The predicted flux for model J of 2.8\,$\times$\,10$^{-5}$\,ph\,cm$^{-2}$\,s$^{-1}$ is just below the upper limit in the 60-100 keV band.
 
In models G-J of \citet{pittard-DSA}, it is considered that the Razin effect is the cause of the turndown of the radio spectrum at GHz frequencies. The fact that these models are rejected -- considering our upper limits -- invalidates this latter scenario, and favors a scenario where the turndown is due to the free-free absorption by the circumstellar winds (models A-F). This model selection implies also that the B-field at the apex of the wind collision region is of order 1~G at the orbital phase examined (0.837) by \citet{pittard-DSA}, rather than a factor of 10 lower (models A-F consider larger magnetic energy densities than models G-J). To place more stringent constraints on the B-field will require actual detection of IC emission. Unfortunately, the predicted IC fluxes from models A-F are typically two orders of magnitude below the upper limits presented here. They are therefore completely our of reach of {\it INTEGRAL} even considering several tens of Ms of observation. 

We note that this comparison between predictions and observations is based on the assumption that the relativistic electron properties at the time of the radio observation considered by \citet{pittard-DSA} are similar at the orbital phases covered by the {\it INTEGRAL} observations. However, as we are dealing with an eccentric binary, the properties of the colliding-wind region are expected to vary with the orbital phase. In order to lift this assumption, simultaneous radio and improved sensitivity X-ray observations are needed.\\

We also investigated the case of the two very long period colliding-wind binaries WR\,146 (WC5 + O8) and WR\,147 (WN8h + B0.5V) (see respectively \citealt{dougherty-wr146} and \citealt{williams-wr147}). Their observational upper limits are the same as for 3EG\,J2033+4118, suggesting that the background reaches a rather uniform level in the region of the image where these targets are located. In the case of WR\,147, we note that much better radio observations are needed in order to apply correctly a model such as that of \citet{pittard-DSA} and make detailed predictions on their non-thermal high-energy emission level to be confronted to our upper limits. Using preliminary fits of radio data for WR\,146 and WR\,147, we predict high-energy fluxes at least one order of magnitude below the observational upper limits. But it is worth noting that the uncertainty on the power law index of the relativistic electron population derived from the fit of radio data is too large to lead too any firm conclusion. We note also that the non detection of WR\,147 is in agreement with the prediction by \citet{reimer} who argued that several Ms of observation with {\it INTEGRAL} may be needed to detect it.

We note also that these three WR systems were considered in the study of \citet{benaglia-romero} of the $\gamma$-ray emission from WR binaries. According to their model, IC fluxes of about 8\,$\times$\,10$^{-4}$, 1\,$\times$\,10$^{-4}$ and 1\,$\times$\,10$^{-3}$\,ph\,cm$^{-2}$\,s$^{-1}$ are expected respectively for WR\,140, WR\,146 and WR\,147 in the {\it INTEGRAL}-IBIS energy range, i.e. between 15\,keV and 10\,MeV. Using the same power law model, we estimated the photon fluxes in the three energy bands used for our data analysis. We obtain predicted values that are larger than our observational upper limits for WR\,140 and WR\,147. As a result, our upper limits for WR\,140 and WR\,147 are inconsistent with the model of \citet{benaglia-romero}.

\subsection{TeV\,2032+4130}

According to \citet{TeV2}, the angular extent of this target is less than 6 arcmin and it can therefore be considered as a point source (the IBIS point spread function is about 12 arcmin). The upper limits on the flux that we derived for the TeV source are the same as those of 3EG\,J2033+4118 in the three energy bands. This is not surprising as the background seems rather homogeneous in the sky region where these two sources -- along with WR\,146 and WR\,147 -- are located (see Fig.\,\ref{mosazoom}). The improved upper limits on the hard X-ray flux from this unidentified object are expected to constitute helpful constraints for future studies aiming at unveiling the nature of recently discovered very-high energy sources. As the process responsible for the high-energy emission is still a completely open issue, we did not make any hypothesis on the emission model to convert the upper limits into energy or photon fluxes. We note that the recent analysis by \citet{butt-tev} of public ISGRI data (including PCFOV data, in addition to the FCFOV data set used in this study) below 300\,keV did not lead to a detection of TeV\,2032+4130 neither.

\subsection{Other sources}
As mentioned in Table\,\ref{initcat}, several point sources have been detected with our data. It is not the purpose of this paper to go through the details for these sources. Most of them are rather bright sources that have been discussed in several papers (see references in Table\,\ref{initcat}), except for the dwarf nova SS\,Cyg. This target is interesting in the sense that not so many dwarf novae have been detected in hard X-rays. The detection of this cataclysmic variable with {\it INTEGRAL} has already been reported in the soft ISGRI bandpass with a detection significance of 7\,$\sigma$ and a count rate of 0.71\,$\pm$\,0.10\,cts\,s$^{-1}$ between 20 and 40\,keV \citep{softcata-isgri}, but SS\,Cyg was not detected at higher energies. The count rate we report here between 20 and 60 keV is 0.77\,$\pm$\,0.06\,cts\,s$^{-1}$ with a detection significance of 12\,$\sigma$. The 3-$\sigma$ upper limits we derived in the 60--100\,keV and 100--1000\,keV energy bands are respectively 0.11 and 0.14\,cts\,s$^{-1}$. In order to compare the count rates obtained by \citet{softcata-isgri}, we ran the {\sc osa} software in narrower energy bands, i.e. 20-40\,keV and 40-60\,keV, and we derived a count rate of 0.69\,$\pm$\,0.05\,cts\,s$^{-1}$ in the former band, with a detection significance of 13\,$\sigma$. SS\,Cyg is still not detected at the 3\,$\sigma$ level above 40\,keV.

\section{Conclusions}

Our investigation of the high energy emission in the ISGRI bandpass did not lead to the detection of point sources related to non-thermal radio emitting massive stars in the Cyg\,OB2 region. We however derived upper limits on the count rate of these sources in three energy bands in order to constrain the high energy emission level of these targets.

These upper limits provide significant constraints on modelling efforts related to the non-thermal emission processes likely to be at work in massive star environments. Several model fits to the synchrotron radio emission from WR140 yield IC fluxes which are inconsistent with the {\it INTEGRAL} upper limits presented here, although the use of isotropic formulae for the IC emission means that there is some uncertainty attached to these predictions (see \citealt{reimer}, and \citealt{pittard-DSA}). A key theme of the \citet{pittard-DSA} models which remain viable is that the non-thermal electron energy spectrum is flatter than the canonical value expected from DSA. It is possible that such a spectrum may be achieved by the further accleration of particles within a highly turbulent wind-wind collision region, as arises when the stellar winds are structured \citep{pittard-clumping}. The spectral slope of any future detected IC emission will allow much tighter constraints to be placed on the underlying population of relativistic particles in these systems and the physical mechanisms behind their acceleration. The advent of future high-energy observatories such as {\it GLAST} in $\gamma$-rays, and {\it NeXT} or {\it SYMBOL-X} in the hard X-ray domain, is expected to open up new prospects in this context.

\begin{acknowledgements}
We are grateful to Alain Detal for the installation of the OSA software. Our thanks go also to Masha Chernyakova and Roland Walter at ISDC for their helpful support, and to Jean-Christophe Leyder for useful discussions. The Li\`ege team acknowledges support from XMM-Newton and INTEGRAL PRODEX contracts (Belspo). JMP thanks the Royal Society for funding. GER was supported by the Grants FIP 5375 (CONICET) and PICT 03-13291 BID 1728/OC-AR (ANPCyT).

\end{acknowledgements}

\bibliographystyle{aa}


\begin{thebibliography}{39}
\expandafter\ifx\csname natexlab\endcsname\relax\def\natexlab#1{#1}\fi

\bibitem[{{Aharonian} {et~al.}(2005){Aharonian}, {Akhperjanian}, {Beilicke},
  {Bernl{\"o}hr}, {B{\"o}rst}, {Bojahr}, {Bolz}, {Coarasa}, {Contreras},
  {Cortina}, {Denninghoff}, {Fonseca}, {Girma}, {G{\"o}tting}, {Heinzelmann},
  {Hermann}, {Heusler}, {Hofmann}, {Horns}, {Jung}, {Kankanyan}, {Kestel},
  {Kohnle}, {Konopelko}, {Kranich}, {Lampeitl}, {Lopez}, {Lorenz}, {Lucarelli},
  {Mang}, {Mazin}, {Meyer}, {Mirzoyan}, {Moralejo}, {O{\~n}a-Wilhelmi},
  {Panter}, {Plyasheshnikov}, {P{\"u}hlhofer}, {de los Reyes}, {Rhode},
  {Ripken}, {Rowell}, {Sahakian}, {Samorski}, {Schilling}, {Siems},
  {Sobzynska}, {Stamm}, {Tluczykont}, {Vitale}, {V{\"o}lk}, {Wiedner}, \&
  {Wittek}}]{TeV}
{Aharonian}, F., {Akhperjanian}, A., {Beilicke}, M., {et~al.} 2005, \aap, 431,
  197

\bibitem[{{Bassani} {et~al.}(2006){Bassani}, {Molina}, {Malizia}, {Stephen},
  {Bird}, {Bazzano}, {B{\'e}langer}, {Dean}, {De Rosa}, {Laurent}, {Lebrun},
  {Ubertini}, \& {Walter}}]{bassani-qso}
{Bassani}, L., {Molina}, M., {Malizia}, A., {et~al.} 2006, \apjl, 636, L65

\bibitem[{{Bazzano} {et~al.}(2003){Bazzano}, {Bird}, {Capitanio}, {Del Santo},
  {Ubertini}, {Zdziarski}, {Di Cocco}, {Falanga}, {Goldoni}, {Goldwurm},
  {Laurent}, {Lebrun}, {Malaguti}, \& {Segreto}}]{bazzano-cygX1}
{Bazzano}, A., {Bird}, A.~J., {Capitanio}, F., {et~al.} 2003, \aap, 411, L389

\bibitem[{{Benaglia} \& {Romero}(2003)}]{benaglia-romero}
{Benaglia}, P. \& {Romero}, G.~E. 2003, \aap, 399, 1121

\bibitem[{{Benaglia} {et~al.}(2001){Benaglia}, {Romero}, {Stevens}, \&
  {Torres}}]{benaglia-5}
{Benaglia}, P., {Romero}, G.~E., {Stevens}, I.~R., \& {Torres}, D.~F. 2001,
  \aap, 366, 605

\bibitem[{{Bird} {et~al.}(2004){Bird}, {Barlow}, {Bassani}, {Bazzano},
  {Bodaghee}, {Capitanio}, {Cocchi}, {Del Santo}, {Dean}, {Hill}, {Lebrun},
  {Malaguti}, {Malizia}, {Much}, {Shaw}, {Stephen}, {Terrier}, {Ubertini}, \&
  {Walter}}]{softcata-isgri}
{Bird}, A.~J., {Barlow}, E.~J., {Bassani}, L., {et~al.} 2004, \apjl, 607, L33

\bibitem[{{Blomme}(2005)}]{blomme-jenam}
{Blomme}, R. 2005, in Massive Stars and High-Energy Emission in OB
  Associations, ed. G.~{Rauw}, Y.~{Naz{\'e}}, \& E.~R. {Blomme}, Gosset, 45--48

\bibitem[{{Butt} {et~al.}(2003){Butt}, {Benaglia}, {Combi}, {Corcoran}, {Dame},
  {Drake}, {Kaufman Bernad{\'o}}, {Milne}, {Miniati}, {Pohl}, {Reimer},
  {Romero}, \& {Rupen}}]{buttcwe-tev}
{Butt}, Y.~M., {Benaglia}, P., {Combi}, J.~A., {et~al.} 2003, \apj, 597, 494

\bibitem[{{Butt} {et~al.}(2007){Butt}, {Combi}, {Drake}, {Finley}, {Konopelko},
  {Lister}, \& {Rodriguez}}]{butt-tev}
{Butt}, Y.~M., {Combi}, J.~A., {Drake}, J., {et~al.} 2007, ArXiv Astrophysics
  e-prints

\bibitem[{{Camero Arranz} {et~al.}(2005){Camero Arranz}, {Wilson}, {Connell},
  {Mart{\'{\i}}nez N{\'u}{\~n}ez}, {Blay}, {Beckmann}, \&
  {Reglero}}]{camero-exo}
{Camero Arranz}, A., {Wilson}, C.~A., {Connell}, P., {et~al.} 2005, \aap, 441,
  261

\bibitem[{{Contreras} {et~al.}(1997){Contreras}, {Rodriguez}, {Tapia},
  {Cardini}, {Emanuele}, {Badiali}, \& {Persi}}]{contreras-cyg5}
{Contreras}, M.~E., {Rodriguez}, L.~F., {Tapia}, M., {et~al.} 1997, \apjl, 488,
  L153+

\bibitem[{{Courvoisier} {et~al.}(2003){Courvoisier}, {Walter}, {Beckmann},
  {Dean}, {Dubath}, {Hudec}, {Kretschmar}, {Mereghetti}, {Montmerle},
  {Mowlavi}, {Paltani}, {Preite Martinez}, {Produit}, {Staubert}, {Strong},
  {Swings}, {Westergaard}, {White}, {Winkler}, \& {Zdziarski}}]{isdc}
{Courvoisier}, T.~J.-L., {Walter}, R., {Beckmann}, V., {et~al.} 2003, \aap,
  411, L53

\bibitem[{{De Becker}(2005)}]{debecker-thesis}
{De Becker}, M. 2005, PhD thesis, University of Li\`ege

\bibitem[{{De Becker} {et~al.}(2005{\natexlab{a}}){De Becker}, {Rauw},
  {Blomme}, {Pittard}, {Stevens}, \& {Runacres}}]{debecker-167971}
{De Becker}, M., {Rauw}, G., {Blomme}, R., {et~al.} 2005{\natexlab{a}}, \aap,
  437, 1029

\bibitem[{{De Becker} {et~al.}(2004{\natexlab{a}}){De Becker}, {Rauw},
  {Blomme}, {Waldron}, {Sana}, {Pittard}, {Eenens}, {Stevens}, {Runacres}, {Van
  Loo}, \& {Pollock}}]{debecker-168112}
{De Becker}, M., {Rauw}, G., {Blomme}, R., {et~al.} 2004{\natexlab{a}}, \aap,
  420, 1061

\bibitem[{{De Becker} {et~al.}(2004{\natexlab{b}}){De Becker}, {Rauw}, \&
  {Manfroid}}]{debecker-cyg8asol}
{De Becker}, M., {Rauw}, G., \& {Manfroid}, J. 2004{\natexlab{b}}, \aap, 424,
  L39

\bibitem[{{De Becker} {et~al.}(2006){De Becker}, {Rauw}, {Sana}, {Pollock},
  {Pittard}, {Blomme}, {Stevens}, \& {van Loo}}]{debecker-cyg8a}
{De Becker}, M., {Rauw}, G., {Sana}, H., {et~al.} 2006, \mnras, 371, 1280

\bibitem[{{De Becker} {et~al.}(2005{\natexlab{b}}){De Becker}, {Rauw}, \&
  {Swings}}]{debecker-hk}
{De Becker}, M., {Rauw}, G., \& {Swings}, J.-P. 2005{\natexlab{b}}, \apss, 297,
  291

\bibitem[{{Dougherty} {et~al.}(2000){Dougherty}, {Williams}, \&
  {Pollacco}}]{dougherty-wr146}
{Dougherty}, S.~M., {Williams}, P.~M., \& {Pollacco}, D.~L. 2000, \mnras, 316,
  143

\bibitem[{{Goldoni} {et~al.}(2003){Goldoni}, {Bonnet-Bidaud}, {Falanga}, \&
  {Goldwurm}}]{goldoni-cygX3}
{Goldoni}, P., {Bonnet-Bidaud}, J.~M., {Falanga}, M., \& {Goldwurm}, A. 2003,
  \aap, 411, L399

\bibitem[{{Hartman} {et~al.}(1999){Hartman}, {Bertsch}, {Bloom}, {Chen},
  {Deines-Jones}, {Esposito}, {Fichtel}, {Friedlander}, {Hunter}, {McDonald},
  {Sreekumar}, {Thompson}, {Jones}, {Lin}, {Michelson}, {Nolan}, {Tompkins},
  {Kanbach}, {Mayer-Hasselwander}, {M{\"u}cke}, {Pohl}, {Reimer}, {Kniffen},
  {Schneid}, {von Montigny}, {Mukherjee}, \& {Dingus}}]{egretcat}
{Hartman}, R.~C., {Bertsch}, D.~L., {Bloom}, S.~D., {et~al.} 1999, \apjs, 123,
  79

\bibitem[{{Kn{\"o}dlseder}(2000)}]{knodlseder}
{Kn{\"o}dlseder}, J. 2000, \aap, 360, 539

\bibitem[{{Konopelko} {et~al.}(2007){Konopelko}, {Atkins}, {Blaylock},
  {Buckley}, {Butt}, {Carter-Lewis}, {Celik}, {Cogan}, {Chow}, {Cui},
  {Dowdall}, {Ergin}, {Falcone}, {Fegan}, {Fegan}, {Finley}, {Fortin},
  {Gillanders}, {Gutierrez}, {Hall}, {Hanna}, {Horan}, {Hughes}, {Humensky},
  {Imran}, {Jung}, {Kaaret}, {Kenny}, {Kertzman}, {Kieda}, {Kildea}, {Knapp},
  {Kosack}, {Krawczynski}, {Krennrich}, {Lang}, {LeBohec}, {Moriarty},
  {Mukherjee}, {Nagai}, {Ong}, {Perkins}, {Pohl}, {Ragan}, {Reynolds}, {Rose},
  {Sembroski}, {Schr{\"o}dter}, {Smith}, {Steele}, {Syson}, {Swordy}, {Toner},
  {Valcarcel}, {Vassiliev}, {Wagner}, {Wakely}, {Weekes}, {White}, {Williams},
  \& {Zitzer}}]{TeV2}
{Konopelko}, A., {Atkins}, R.~W., {Blaylock}, G., {et~al.} 2007, \apj, 658,
  1062

\bibitem[{{Kuznetsov} {et~al.}(2003){Kuznetsov}, {Falanga}, {Blay}, {Goldwurm},
  {Goldoni}, \& {Reglero}}]{kuznetsov-exo}
{Kuznetsov}, S., {Falanga}, M., {Blay}, P., {et~al.} 2003, \aap, 411, L437

\bibitem[{{Lebrun} {et~al.}(2003){Lebrun}, {Leray}, {Lavocat}, {Cr{\'e}tolle},
  {Arqu{\`e}s}, {Blondel}, {Bonnin}, {Bou{\`e}re}, {Cara}, {Chaleil}, {Daly},
  {Desages}, {Dzitko}, {Horeau}, {Laurent}, {Limousin}, {Mathy}, {Mauguen},
  {Meignier}, {Molini{\'e}}, {Poindron}, {Rouger}, {Sauvageon}, \&
  {Tourrette}}]{lebrun-isgri}
{Lebrun}, F., {Leray}, J.~P., {Lavocat}, P., {et~al.} 2003, \aap, 411, L141

\bibitem[{{Marchenko} {et~al.}(2003){Marchenko}, {Moffat}, {Ballereau},
  {Chauville}, {Zorec}, {Hill}, {Annuk}, {Corral}, {Demers}, {Eenens}, {Panov},
  {Seggewiss}, {Thomson}, \& {Villar-Sbaffi}}]{marchenko-wr140}
{Marchenko}, S.~V., {Moffat}, A.~F.~J., {Ballereau}, D., {et~al.} 2003, \apj,
  596, 1295

\bibitem[{{Natalucci} {et~al.}(2003){Natalucci}, {Del Santo}, {Ubertini},
  {Capitanio}, {Cocchi}, {Piraino}, \& {Santangelo}}]{natalucci-cygX2}
{Natalucci}, L., {Del Santo}, M., {Ubertini}, P., {et~al.} 2003, \aap, 411,
  L395

\bibitem[{{Pittard}(2007)}]{pittard-clumping}
{Pittard}, J.~M. 2007, \apj, 660, L141

\bibitem[{{Pittard} \& {Dougherty}(2006)}]{pittard-DSA}
{Pittard}, J.~M. \& {Dougherty}, S.~M. 2006, \mnras, 372, 801

\bibitem[{{Rauw} {et~al.}(2002){Rauw}, {Blomme}, {Waldron}, {Corcoran},
  {Pittard}, {Pollock}, {Runacres}, {Sana}, {Stevens}, \& {Van
  Loo}}]{rauw-9sgr}
{Rauw}, G., {Blomme}, R., {Waldron}, W.~L., {et~al.} 2002, \aap, 394, 993

\bibitem[{{Reimer} {et~al.}(2006){Reimer}, {Pohl}, \& {Reimer}}]{reimer}
{Reimer}, A., {Pohl}, M., \& {Reimer}, O. 2006, \apj, 644, 1118

\bibitem[{{Romero} {et~al.}(1999){Romero}, {Benaglia}, \&
  {Torres}}]{romero-egret}
{Romero}, G.~E., {Benaglia}, P., \& {Torres}, D.~F. 1999, \aap, 348, 868

\bibitem[{{Sidoli} {et~al.}(2005){Sidoli}, {Mereghetti}, {Larsson},
  {Chernyakova}, {Kreykenbohm}, {Kretschmar}, {Paizis}, {Santangelo},
  {Ferrigno}, \& {Falanga}}]{sidoli-sax}
{Sidoli}, L., {Mereghetti}, S., {Larsson}, S., {et~al.} 2005, \aap, 440, 1033

\bibitem[{{Torres} {et~al.}(2004){Torres}, {Domingo-Santamar{\'{\i}}a}, \&
  {Romero}}]{torres-assoc}
{Torres}, D.~F., {Domingo-Santamar{\'{\i}}a}, E., \& {Romero}, G.~E. 2004,
  \apjl, 601, L75

\bibitem[{{Tsygankov} \& {Lutovinov}(2005)}]{tsygankov-ks}
{Tsygankov}, S.~S. \& {Lutovinov}, A.~A. 2005, Astronomy Letters, 31, 88

\bibitem[{{Ubertini} {et~al.}(2003){Ubertini}, {Lebrun}, {Di Cocco}, {Bazzano},
  {Bird}, {Broenstad}, {Goldwurm}, {La Rosa}, {Labanti}, {Laurent}, {Mirabel},
  {Quadrini}, {Ramsey}, {Reglero}, {Sabau}, {Sacco}, {Staubert}, {Vigroux},
  {Weisskopf}, \& {Zdziarski}}]{ibis}
{Ubertini}, P., {Lebrun}, F., {Di Cocco}, G., {et~al.} 2003, \aap, 411, L131

\bibitem[{{Van Loo}(2005)}]{vanloo-thesis}
{Van Loo}, S. 2005, PhD thesis, University of Leuven

\bibitem[{{Walter} {et~al.}(2004){Walter}, {Bodaghee}, {Barlow}, {Bird},
  {Dean}, {Hill}, {Shaw}, {Bazzano}, {Ubertini}, {Bassani}, {Malizia},
  {Stephen}, {Belanger}, {Lebrun}, \& {Terrier}}]{walter-igr}
{Walter}, R., {Bodaghee}, A., {Barlow}, E.~J., {et~al.} 2004, The Astronomer's
  Telegram, 229, 1

\bibitem[{{Williams} {et~al.}(1997){Williams}, {Dougherty}, {Davis}, {van der
  Hucht}, {Bode}, \& {Setia Gunawan}}]{williams-wr147}
{Williams}, P.~M., {Dougherty}, S.~M., {Davis}, R.~J., {et~al.} 1997, \mnras,
  289, 10

\end{thebibliography}

\end{document}